\pdfoutput=1
\documentclass{aa}  
\usepackage{graphicx}
\usepackage{txfonts}
\usepackage{natbib,twoopt}
\usepackage[breaklinks=true]{hyperref} 
\bibpunct{(}{)}{;}{a}{}{,}             
\usepackage{multirow} 

\newcommand{\cloudy}{{\it Cloudy}}
\newcommand{\swift}{{\it Swift}}

\newcommand{\uvmt}{{\it uvm2}}
\newcommand{\wtw}{WASP-121}
\newcommand{\wtwb}{WASP-121\,b}
\newcommand{\wt}{WASP-12}
\newcommand{\wtb}{WASP-12\,b}

\newcommand{\hdtb}{HD\,209458\,b}

\newcommand{\ffc}{55\,Cnc}      
\newcommand{\ffce}{55\,Cnc\,e}  

\newcommand{\funit}{erg\,cm$^{-2}$\,s$^{-1}$}
\newcommand{\ergs}{erg\,s$^{-1}$}

\newcommand{\mlossunit}{g\,s$^{-1}$}
\newcommand{\gccm}{g\,cm$^{-3}$}
\newcommand{\ccm}{cm$^{-3}$}

\newcommand{\cnts}{cts\,s$^{-1}$}
\newcommand{\cntks}{cts\,ks$^{-1}$}

\newcommand{\Rpl}{R$_{\rm p}$}

\newcommand{\hto}{H$_2$O}
\begin{document} 

   \title{{\bfseries\itshape Swift} UVOT near-UV transit observations of WASP-121\,b\thanks{
   Light curves shown in Fig. A.1 are available in electronic
   form at the CDS via anonymous ftp to cdsarc.u-strasbg.fr (130.79.128.5)
   or via http://cdsweb.u-strasbg.fr/cgi-bin/qcat?J/A+A/TBD}}
   
   \titlerunning{{\it Swift} UVOT near-UV transit observations of WASP-121\,b}

   \author{M. Salz\inst{1}, 
           P. C. Schneider\inst{1},
           L. Fossati\inst{2,3},
           S. Czesla\inst{1},
           K. France\inst{4},
           J. H. M. M. Schmitt\inst{1}
          }
          
   \authorrunning{Salz et al.}

   \institute{Hamburger Sternwarte, Universit\"at Hamburg,
              Gojenbergsweg 112, 21029 Hamburg, Germany\\
              \email{msalz@hs.uni-hamburg.de}
         \and
              Space Research Institute, Austrian Academy of Sciences, 
              Schmiedlstrasse 6, A-8042 Graz, Austria
         \and
              School of Physical Sciences, The Open University, Walton Hall, 
              Milton Keynes MK7 6AA, UK
         \and
              Laboratory for Atmospheric and Space Physics, University of Colorado,
              600 UCB, Boulder, CO 80309, USA
             }

   \date{Received December 5, 2017 / Accepted 23 January 2019}

   \abstract{
Close-in gas planets are subject to continuous photoevaporation that can erode their volatile envelopes. Today, ongoing mass loss has been confirmed in a few individual systems via transit observations in the ultraviolet spectral range. We demonstrate that the Ultraviolet/Optical Telescope (UVOT) onboard the Neil Gehrels {\it Swift} Observatory enables photometry to a relative accuracy of about 0.5\% and present the first near-UV (200\,$-$\,270~nm, NUV) transit observations of \wtwb{}, a hot Jupiter with one of the highest predicted mass-loss rates. 
The data cover the orbital phases 0.85 to 1.15 with three visits.
We measure a broad-band NUV transit depth of $2.10\pm0.29$\%.
While still consistent with the optical value of 1.55\%, the NUV data indicate excess absorption of 0.55\% at a 1.9$\sigma$ level.
Such excess absorption is known from the \wt{} system, and both of these hot Jupiters
are expected to undergo mass loss at extremely high rates.
With a \cloudy{} simulation, we show that absorption lines of \ion{Fe}{ii} in a dense extended atmosphere can cause broad-band NUV absorption at the 0.5\% level.
Given the numerous lines of low-ionization metals, the NUV range is a promising tracer of photoevaporation in the hottest gas planets.
     }


   \maketitle

\section{Introduction}\label{SectIntro}
%
Shortly after its discovery \citep{Henry2000, Charbonneau2000}, the first transiting hot Jupiter \hdtb{} was found to host an extended hydrogen atmosphere that covers approximately 15\% of the host star during transit \citep{Vidal2003}. The atmospheric expansion is thought to be caused by the absorption of high-energy stellar photons (X-rays and extreme UV, XUV, $\lambda<$\,912~\AA{}). The ensuing heat input raises the temperature in planetary thermospheric layers to about $10^4$~K and accelerates atmospheric gas to supersonic speeds. This results in the formation of a planetary wind that somewhat resembles the solar wind \citep{Parker1958}.  Estimates for the resulting planetary mass-loss rates are often obtained by assuming that this planetary wind is energy-limited \citep{Watson1981}. However, the true nature of such winds can be more complex, because simulations have shown a transition from the energy- to a recombination-limited regime \citep{Owen2016, Salz2016b}. Additionally, interactions with the stellar wind or a planetary magnetic field further complicate matters \citep[e.g.,][]{Murray2009, Khodachenko2015}.

Planetary winds can be studied indirectly, because they are thought to shape the radius distribution of short-period planets \citep{Lundkvist2016, Fulton2017}. However, direct observations of evaporating atmospheres in four systems have been crucial for our understanding of atmospheric escape: HD\,209458\,b, HD\,189733\,b, \wtb{}, and GJ\,436\,b \citep{Vidal2003, Lecavelier2010, Linsky2010, Lecavelier2012, Fossati2010, Haswell2012, Kulow2014,Ehrenreich2015,Lavie2017}.

Among the direct detections, \object{WASP-12} is the most extreme system: The planet orbits just above the Roche-lobe limit and experiences one of the highest irradiation levels of any hot Jupiter \citep{Hebb2009}. This results in a powerful planetary wind, as evidenced by Hubble Space Telescope (HST) Cosmic Origins Spectrograph (COS) observations that have revealed  a transit depth of 3\% in three near-UV (NUV) bands between 254 and 283~nm. The NUV transit is about two times deeper than the optical one of 1.4\%
\citep{Fossati2010, Haswell2012, Nichols2015}. This planet is likely surrounded by a thick, escaping atmosphere in which metals are dragged into the upper atmosphere and low-ionization metallic absorption lines cause the additional broad-band NUV absorption \citep{Fossati2010}.

\section{The WASP-121 system}
%
With respect to the expected planetary mass loss, \object{WASP-121} is a close analog to the \wt{} system \citep{Delrez2016}, but the host star is four times brighter. A hot Jupiter orbits the bright F6-type main-sequence star (${\rm V} =10.4$, $T_{\rm eff}=6460$~K) in 1.27 days, close to the tidal disruption limit \citep[1.15\,$\times$ Roche limit,][]{Delrez2016}. \wtwb{} is subject to an exceptionally high bolometric insolation of $7.1\times 10^9$~\funit{}, 5200 times that of the Earth. It exhibits a mean density of only 0.183~\gccm{} and in Sect.~\ref{SectDiscussion} we further derive a high level of XUV irradiation. This should lead to severe mass loss, since the evaporation mainly depends on the irradiation level and the mean planetary density \citep[e.g.,][]{Erkaev2007}.

Activity of the host star manifests itself in an elevated level of radial-velocity (RV) jitter, which has provided an estimate for the stellar rotation period of $\approx$\,1~d  \citep{Delrez2016}. The star is likely seen at a low inclination about 10\degr{} with the planet in a nearly polar orbit. The lack of \ion{Ca}{ii} H and K emission line cores in a presumably active star is reminiscent of the \wt{} system, where material evaporated from the hot Jupiter is thought to veil stellar chromospheric emission \citep{Fossati2013}.

HST Wide Field Camera~3 observations of the primary and secondary transits have revealed \hto{} and  possibly indicate TiO, VO, and FeH in the planetary atmosphere \citep{Evans2016, Tsiaras2017}. \hto{} emission in the secondary transit provides the first clear evidence of a stratospheric temperature inversion in a hot Jupiter \citep{Evans2017}.

\section{Photometric stability of \textit{Swift}'s UVOT}\label{Sect:stability}
%
Ultraviolet transit observations with the HST have revealed four examples of escaping atmospheres, but HST resources are limited. The Ultraviolet/Optical Telescope (UVOT) onboard the Neil Gehrels {\it Swift} Observatory also provides an opportunity to obtain space-based UV photometry, albeit with a smaller telescope \citep[30~cm primary mirror compared to HST's 2.4~m mirror,][]{Roming2005, Poole2008, Breeveld2010}. The instrument is a micro-channel plate intensified photon counting detector. In event mode each photon is registered with the timing resolution of the CCD read out (full frame $\approx$\,11~ms). \swift{} is in a low Earth orbit with a period of 85~min, allowing for undisturbed visibility windows of typically about 30~min -- a little shorter than the visibility windows of the HST. Hence, transit observations that last a few hours must extend over several spacecraft orbits.

To test the stability of the UVOT and determine its suitability for transit observations, we analyzed an archival data set of \ffce{} (observation IDs 00034849001 to 00034849012). The host star was observed in three visits over four consecutive spacecraft orbits and in one visit over five orbits (24 Dec. 2016, 18 Mar. 2017, 19 Mar. 2017, and 12 May 2017). The chosen \uvmt{} filter has a central wavelength of 225~nm and a full width at half maximum (FWHM) of 50~nm.
To achieve the best photometric accuracy, the target was placed onto similar positions on the detector, which requires two correction slews. Thus, prior to the about 1500~s long science exposure, each orbit contains two acquisition exposures of about 200~s duration. The final light curves are affected by known regions of low sensitivity, which are likely caused by dust depositions on the photocathode. Since no corrections are available, we excluded two affected exposures (see the procedure in the Appendix and SWIFT-UVOT-CALDB-17-01b\footnote{\href{https://heasarc.gsfc.nasa.gov/docs/heasarc/caldb/swift/docs/uvot/}{https://heasarc.gsfc.nasa.gov/docs/heasarc/caldb/swift/docs/uvot/}}). 

We reduced the data using HEASOFT (v6.20). The tasks {\tt coordinator} and {\tt uvotscreen} were run with standard input parameters to obtain an event table in sky coordinates, filtered for good time intervals and data quality. The attitude control during these observations was inaccurate, which led to a rejection of large chunks of data unless photons with the imperfect attitude flag 256 were accepted (private com. P. Kuin).

To account for the residual motion of the source caused by the imperfect attitude control, we devised a nonstandard reduction procedure in collaboration with the UVOT team to increase the photometric stability. We used {\tt uvotscreen} to slice the event tables into 20~s bins, determined the source positions in each slice via iteratively finding its centroid in ds9, and then ran the standard {\tt uvotevtlc} routine on each slice to apply the corrections for background, aperture, coincidence loss, dead time, large-scale sensitivity variations, and long-term sensitivity variations. The light curve slices were merged with {\tt ftmerge} and the time stamps were converted to the barycentric dynamical time (TDB) scale via {\tt barycorr}.

\begin{figure}[tb]
  \centering
  \includegraphics[draft=false, width=\hsize, trim = 2mm 2mm 2mm 2mm, clip]{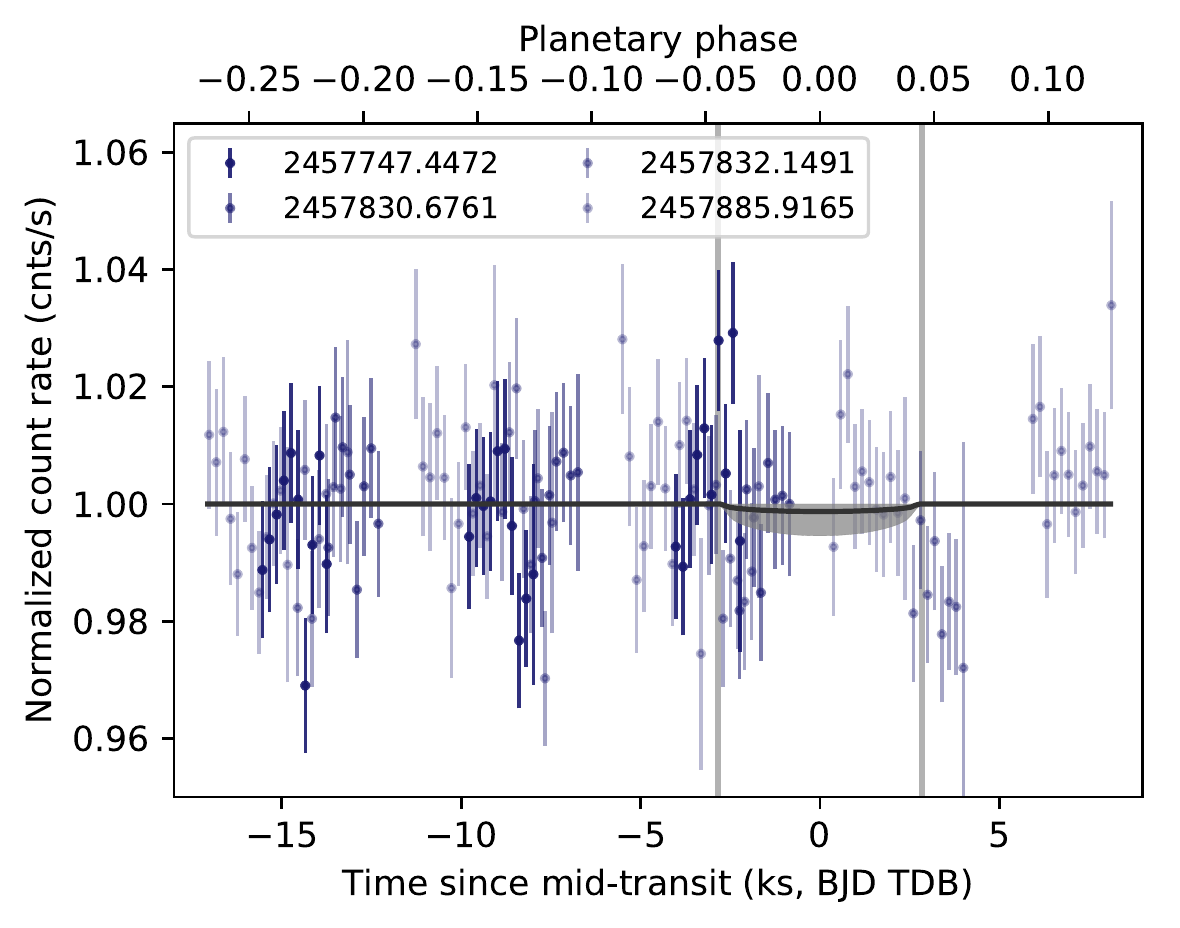}
  \caption{\swift{} UVOT  transit light curves of \ffce{} binned to 200~s.
           Individual visits are color coded; mid-transit times are given in
           the legend.
           The planetary transit is marked by vertical lines.
           The best fit Mandel-Agol transit model (black line) is shown with the
           95\% confidence region (gray shaded area): 
           A planetary transit is not detected. The four light curves visualize the
           stability of \swift{}'s UVOT.
           }
  \label{fig:55cnc}
\end{figure}

The obtained light curves of \ffc{} are shown in Fig.~\ref{fig:55cnc}.The hot super-Earth has a visual transit depth of 0.04\%, which is not detected in the UVOT data. With a count rate of 117~\cnts{}, the data are considerably affected by coincidence loss, which is however well controlled \citep{Poole2008, Kuin2008}. The coincidence loss affects the standard deviation during the individual orbits, which is a factor 1.5 higher than the photon noise.

During each visit, we detect variations with a standard deviation of 0.42\% of the average source rates from orbit to orbit. This instrumental error is likely caused by small-scale sensitivity variations of the detector and thus depends on the source position on the detector. The source position is found to be random in the vicinity of the aimpoint after the two correction slews, and therefore this instrumental error can be assumed to be random. 
Thus, \swift{} UVOT achieves an accuracy in relative flux measurements of about 0.42\% over several orbits and is suited to measure enhanced NUV transit depths of around 3\% as observed in \wtb{}. However, multiple in- and out-of-transit exposures are required to achieve sufficient accuracy, which usually requires  transit observations to be repeated in several visits.

\begin{figure*}[tb]
  \centering
  \begin{minipage}[t]{.62\textwidth}
    \vspace{0pt}
    \includegraphics[draft=false, width=\hsize, trim = 2.5mm 2.5mm 3.5mm 2.5mm, clip]{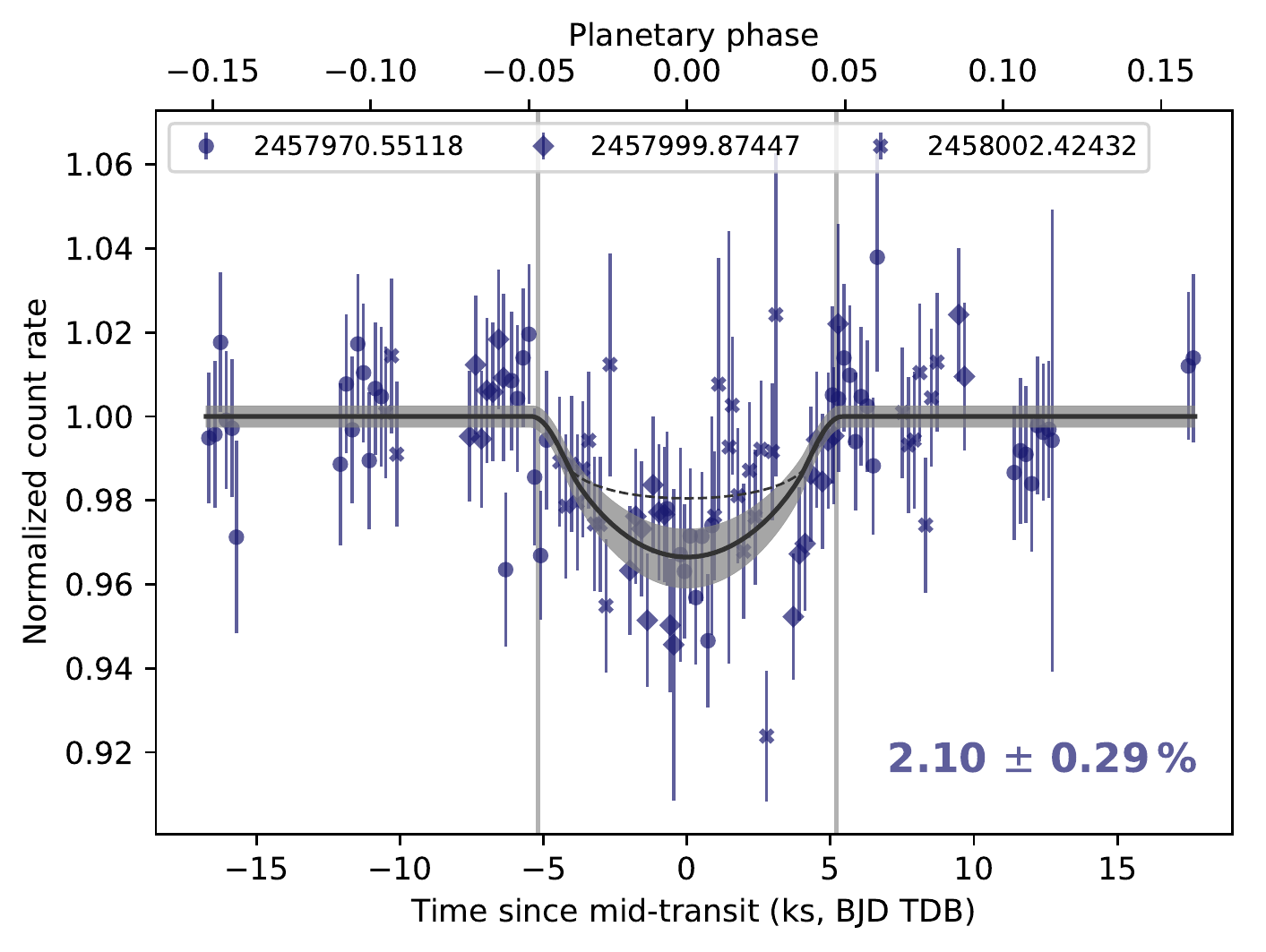}
  \end{minipage}
  \hfill
  \begin{minipage}[t]{0.375\textwidth}
    \vspace{7.7mm}
    \includegraphics[draft=false, width=\hsize, trim = 1.2mm 2.5mm 3.5mm 3mm, clip]{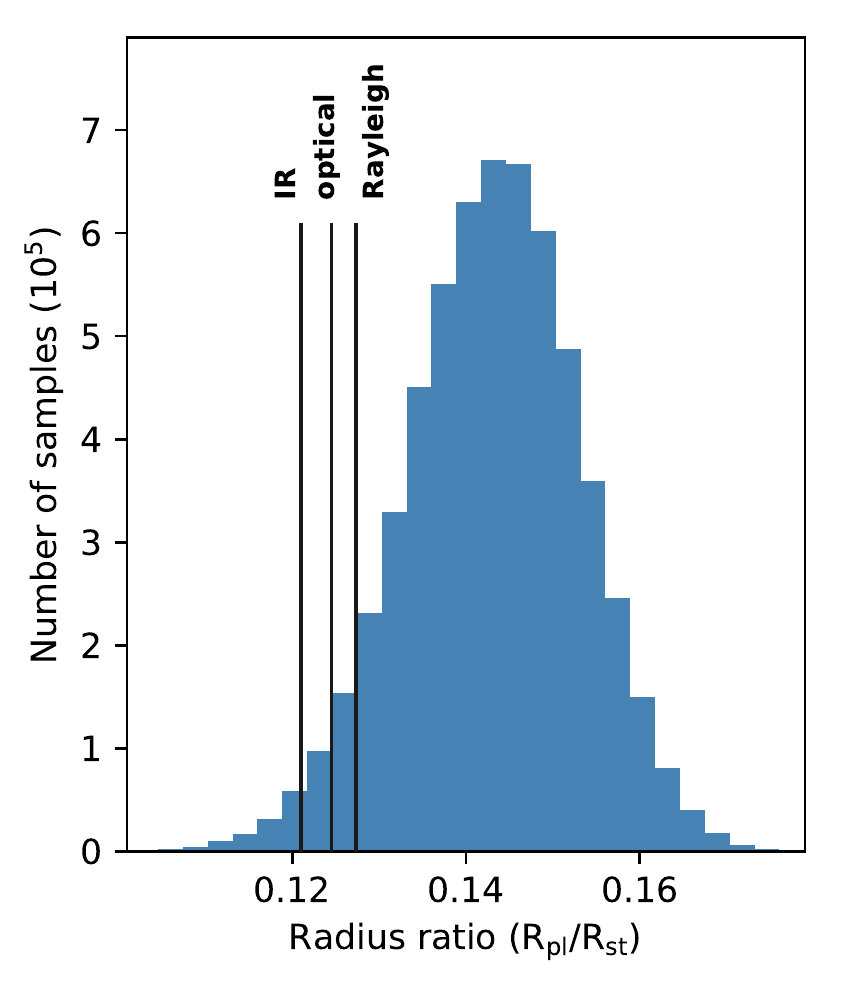}
  \end{minipage}
  \caption{{\it Left:} Phased light curves of WASP-121 during three transits observed with
           \swift{} UVOT labeled by their mid-transit times in BJD.
           The data are rebinned to 200~s intervals and corrected for the
           modeled systematic offsets. The uncorrected data are shown in
           Fig.~\ref{fig:discard}.
           The transit model with 1$\sigma$
           uncertainty interval of the out-of-transit flux level and transit depth is shown. The optical transit
           depth is depicted by the dashed line \citep[B-filter,][]{Delrez2016}. The observed transit depth is
           given at the bottom with 1$\sigma$ error.
           {\it Right:}
           Posterior of the radius ratio.
           Optical and IR values are indicated 
           \citep{Delrez2016, Evans2016} together with the expected NUV transit depth
           according to Rayleigh scattering.
           }
  \label{fig:lc}
\end{figure*}

\section{Transit analysis on the example of 55 Cnc}\label{Sect:55cnc}
%
We analyzed the \ffce{} data implementing the analytical transit model for a circular orbit from \citet{Mandel2002}. The model was explored with a Markov Chain Monte Carlo (MCMC) method following the {\tt emcee} implementation \citep{ForemanMackey2013} and the PyAstronomy package\footnote{\href{https://github.com/sczesla/PyAstronomy}{https://github.com/sczesla/PyAstronomy}}.
The chain was run with the number of independent walkers set to two times the number of free parameters. Each walker had a burn-in of $2\times10^4$ steps and was then propagated over $8\times10^4$ steps, resulting in a total of $5.9\times 10^6$ samples. Besides the radius ratio between the planet and the star, the model has four parameters (semimajor axis, inclination, orbital period, and reference mid-transit time), which are set up according to the values of \citet{Demory2016}. Additionally, we fit the out-of-transit flux level of the host star in each visit individually. 

The instrumental error of the UVOT described in the previous section is introduced into the transit model by allowing multiplicative offsets for each individual exposure with normal priors centered at 1.0 and with a standard deviation of 0.0042. With this procedure, the instrumental error directly affects the marginalized radius ratio and the significance with which a transit is detected.

The data only provide an upper limit for the NUV transit depth of 0.4\% (95\% confidence), compared to the optical transit depth of 0.04\%.

\section{Ultraviolet transit photometry of WASP-121 b}\label{SectRedu}
%
Having validated the capabilities of \swift{} UVOT, we observed three transits of \wtwb{} on Aug 4, Sept 3, and Sept 5, 2017 (observation IDs 00010206001 to 00010206027). As for \ffc{}, the \uvmt{} filter was chosen to avoid the red-leaks present in the other UV filters \citep{Breeveld2011}. The observed flux is dominated by the 200 to 270 nm range, which covers an approximately five times broader wavelength range than the HST COS observations of \wt{}. (see Fig.~\ref{fig:filter}).
Our first visit of \wtw\ extended over seven spacecraft orbits ($\approx$\,35~ks), which allows us to study pre- and post-transit variability; the following two visits extended over four spacecraft orbits. The data were acquired and reduced as described for \ffc{} in Sect.~\ref{Sect:stability}. We identified and excluded five exposures, which were affected by low-sensitivity patches (see labels 1 to 5 in Fig.~\ref{fig:discard} and the discussion in the Appendix. 

The transit of \wtwb\ was analyzed following the description in Sect.~\ref{Sect:55cnc}. System parameters were set up with Gaussian priors according to the values and errors given by \citet{Delrez2016, Evans2016}. These parameters were not further constrained by the UVOT data, but were included as nuisance parameters for their impact on the uncertainty of the transit depth. In total, the model has 37 free parameters: 5 from the transit model, 3 from the out-of-transit flux levels of the three visits, and 29 from the offset parameters for the individual exposures.

The data quality is not sufficient to constrain the coefficients of limb-darkening laws, and for UV filters, they are not available in the literature \citep[e.g.,][]{Claret2011}.
Therefore, we used stellar model atmospheres and angle-dependent specific intensities computed with the PHOENIX code \citep{Husser2013} to derive a theoretical limb-darkening curve for the \uvmt{} filter. We chose stellar parameters that match those of \wtw{} but varying the effective temperature in the range $T_{\rm eff} = 6500\pm200$~K and the surface gravity in the range $\log_{10} g \,{\rm [cm/s]} = 4.0 - 4.5$. 
The quadratic limb-darkening law as used by \citet{Claret2011} with the coefficients $a=1.42\pm0.08$ and $b=-0.49\pm0.08$ fits the theoretical curves of \wtw{}-like stars well\footnote{We followed the rescaling of the $\mu$-axis as done by \citet{Mueller2015}.}. We used the best fit limb-darkening coefficients for the MCMC, but checked that varying the limb-darkening coefficients within twice their derived uncertainty margin has a small impact of less than 0.1 percentage points on the resulting transit depth.

Figure~\ref{fig:lc} shows the NUV light curves phased to the transit of \wtwb{}. The mean out-of-transit count rate of 35.1~\cnts{} varies by nearly 1\% between our visits, indicating slight stellar variability over the observing campaign of one month. The planetary transit is clearly detected with a mean transit depth of $2.10 \pm 0.29$\%, which is 1.4 times deeper than the optical value of $1.551 \pm 0.012$\,\% measured at around 500~nm, but only at at 1.9$\sigma$ level. Rayleigh scattering in a clear atmosphere increases the NUV transit depth at 225~nm to only 1.62\% \citep[using Eq.~1 of][]{Lecavelier2008}. The right-hand side of Fig.~\ref{fig:lc} compares the marginalized radius ratio of our MCMC run with the optical and infrared values. In contrast to the \wt{} system \citep{Fossati2010, Nichols2015}, we do not find significant indications for pre-transit absorption in the \wtw{} data.

\begin{table}
  \caption{MCMC best-fit statistics with 1$\sigma$ error in brackets.}
  \label{TabFit}
  \centering
  \begin{tabular}{l c c}
    \hline\hline\vspace{-7pt}\\
                            & full model & fixed radius ratio \\
    \vspace{-9pt}\\\hline\vspace{-7pt}\\
      radius ratio          & 0.145 (10) & 0.12454 (48) \\
      free parameters       & 37         & 36 \\
      DOF\tablefootmark{a}  & 965        & 966    \\
      $\chi_{\rm red}$      & 1.062      & 1.064    \\
    \vspace{-9pt}\\\hline
  \end{tabular}
  \tablefoot{
             \tablefoottext{a}{degrees of freedom}
            }
\end{table}

Our analysis indicates excess NUV absorption, but the significance remains low. As a reference, we repeated the MCMC transit analysis keeping the radius ratio fixed to the optical value. Unsurprisingly, this model also yields a satisfactory fit to the data (see Fig.~\ref{fig:discard} and Table~\ref{TabFit}). We find that two factors impede a more significant result: uncertainties in the offsets and the out-of-transit flux levels.
In the fixed-radius-ratio model, the in-transit offsets are on average 0.21 percentage points lower than the out-of-transit ones, thus, absorbing about half of the possible excess absorption. It remains unclear whether this is a chance finding, given the relatively small shift of the eight in-transit exposures compared to the assumed standard deviation of 0.42\%.
Additionally, the out-of-transit flux levels for two visits remain more weakly constrained than the in-transit levels. The fixed-radius-ratio model yields estimates for the out-of-transit flux levels on average 0.26\% lower than our standard model (see Fig.~\ref{fig:discard}), again, neutralizing a potentially deeper transit. Therefore, we can neither rule out nor confirm a physical origin of the deeper transit with certainty.

\section{X-ray emission from WASP-121}\label{SectXRT}
%
The \swift{} space observatory provides one advantage compared to HST observations: the X-ray Telescope \citep[XRT,][]{Burrows2005} operates contemporaneously to the UVOT and can provide the planetary high-energy irradiation level at the time of the transit observation.
In our observations, \wtw{} is detected in the first visit with $13.0\pm3.9$ source counts ($1.26\pm0.37$~\cntks{}), but the second and third visits result in a nondetection with an upper limit of $0.3$~\cntks{} (95\% confidence).  Half of the photons in the first visit are detected in the first spacecraft-orbit, indicating flaring activity at the beginning of the observation. Therefore, \wtw{} shows X-ray variability.

\begin{figure}[t]
  \centering
  \includegraphics[draft=false, width=\hsize, trim = 0.5mm 0.5mm 5mm 3mm, clip]{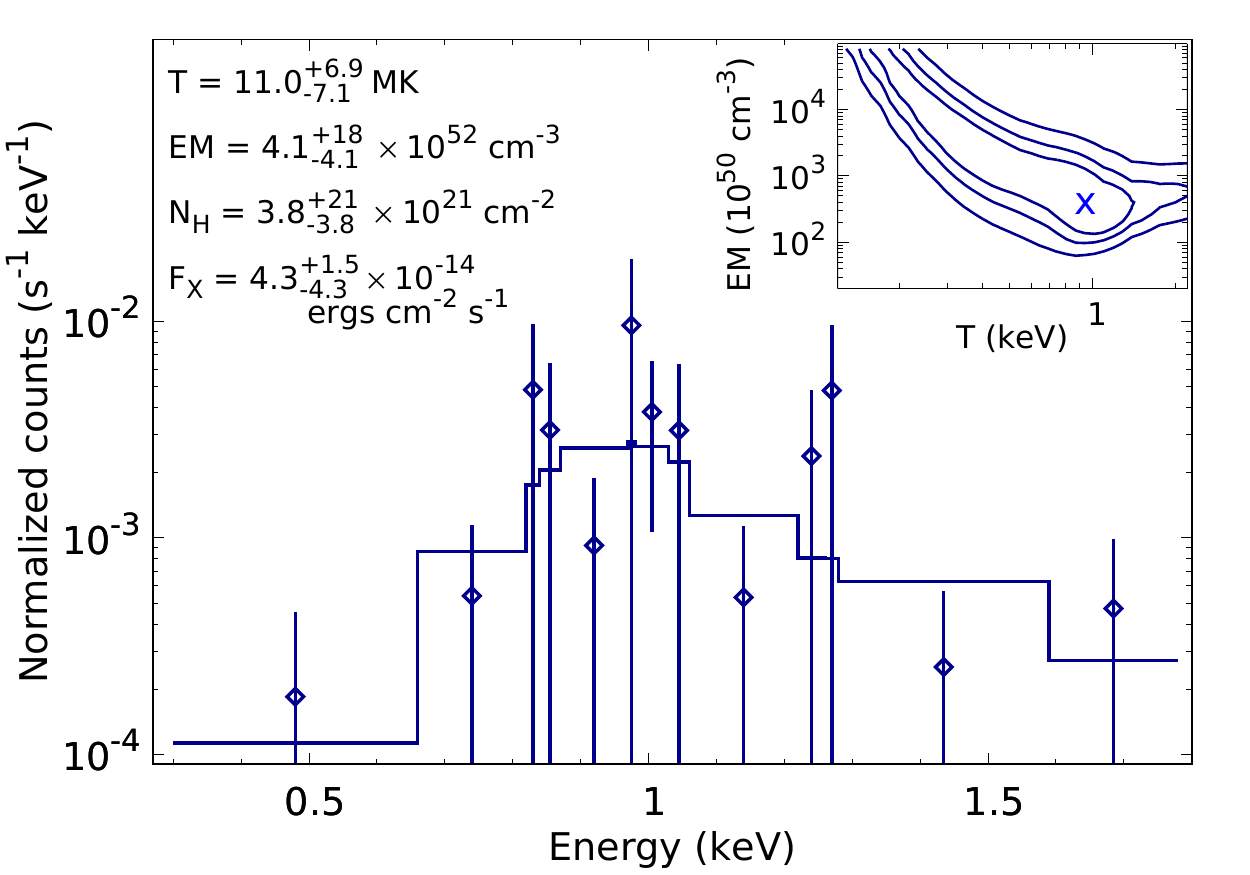}
  \caption{XRT data of the first visit binned to one count per bin.
           The depicted model is a one-temperature fit with an
           interstellar absorption component.
           The insert shows the $\chi^2$ contours of the temperature
           vs. emission measure, derived with
           {\it Xspec}'s {\tt steppar} function.
           The fitted model parameters and the resulting flux
           are provided with 1$\sigma$ uncertainty ranges.
           }
  \label{fig:xrt_spec}
\end{figure} 

We obtained a source spectrum for the first visit from the online interface of the UK Swift Science Data Centre at the University of Leicester \citep[based on HEASOFT v6.20,][]{Evans2009}. The data was modeled with {\it Xspec} (v12.9.1) with an optically thin plasma emission model \citep[APEC, AtomDB v3.0.7,][]{Foster2012} implementing solar abundances \citep{Grevesse1998}. We included an absorption component via the T\"ubingen-Boulder ISM absorption model to fit the equivalent hydrogen column \citep{Wilms2000}. The spectrum of \wtw{} shows hot plasma emission with a peak around 1~keV; the model parameters with 90\% confidence intervals are given in Fig.\ref{fig:xrt_spec}. Given the distance of $272\pm1.6$~pc \citep[][]{Gaia2016A, Gaia2018}, we find an average X-ray luminosity in the first visit of $L_{\rm X}(0.2\!-\!2.4\,\text{keV}) = 1.0^{+3.5}_{-1.0}\times10^{30}$\ergs{}.

For an F-type star with a rotation period of around 1~d, we expect saturated X-ray emission on the order of $10^{30}$\ergs{} \citep{Pizzolato2003}, which is only seen during the supposed flare. The nondetection during visits 2 and 3 implies a more than four times lower average X-ray luminosity or a high column density on the order of $10^{21}$~cm$^{-2}$ that absorbs an intrinsic X-ray luminous source. Such a high column density is consistent with the data of the first visit, where few soft X-rays are detected. This may indicate an elevated level of system intrinsic absorption caused by material that has been evaporated from the planet as suggested for \wt{} \citep{Fossati2013}.

\section{Putting WASP-121 and WASP-12 in context}\label{SectDiscussion}
%
At face value, the NUV absorption of \wtwb{} is on a par with what has been observed by HST in the \wt{} system, demonstrating that such extreme systems may commonly exhibit strong excess broad-band NUV absorption during transit. The planet's Roche lobe cross-section projected onto the star covers 2.5\% of the stellar disk compared to the 2.1\% NUV transit depth. If confirmed, this suggests that the excess NUV absorption occurs in a dense escaping atmosphere that fills the Roche lobe.

\wtw{} is a fast rotator with a 1~d period and is presumably more active than \wt{} with a rotation period estimate of 6~d \citep{Albrecht2012}. These periods provide canonical estimates for the stellar X-ray luminosities of $\log_{10} L_{\rm X}~[{\text{\ergs}}] = 30.0$ and 29.1 for \wtw{} and \wt{} \citep[][]{Pizzolato2003}. Based on the X-ray-EUV flux relations from \citet{King2017}, we derive irradiation levels of $\log_{10} F_{\rm XUV}~[{\text{\funit}}] = 6.2$ and 5.6, which results in energy-limited mass-loss rates of $\log_{10}\dot{M}~[{\text{\mlossunit}}] = 13.7$ and 13.1 for \wtwb{} and \wtb{}\footnote{The derived mass loss of \wtb{} is higher than literature values \citep{Ehrenreich2011, Salz2016}, because we assume faster stellar rotation.} respectively \citep[using the heating efficiency and XUV absorption radius from][]{Salz2016b}. 

Although some uncertainties remain, \wtwb{} and \wtb{} likely experience similarly extreme mass loss on the order of $10^{13}$~\mlossunit. At much lower mass-loss rates on the order of $\approx$\,$10^{10}$~\mlossunit{}, \hdtb{} and WASP-80\,b do not exhibit significantly increased broad-band NUV transit depths \citep{Vidal2013, King2017}. Broad-band NUV absorption could thus serve as a proxy for strong planetary mass loss. This is also supported by simulations, which show that higher mass loss results in higher thermospheric densities \citep[e.g.,][]{Salz2016}.

\section{Can metals cause broad-band NUV absorption?}\label{SectResults}
%
\citet{Fossati2010} and \citet{Haswell2012} have provided evidence that the broad-band NUV absorption in \wtb{} is caused by absorption lines of lowly ionized metals. Our observations do not contain spectral information, and the \uvmt{} filter covers a broader wavelength range than the COS orders. 
We ran a simulation with the \cloudy{} code \citep[C17.00,][]{Ferland2017} to check whether metals in an extended planetary atmosphere can possibly cause the observed absorption level. \cloudy{} self-consistently solves the radiative transfer in an irradiated cloud including the 30 lightest elements. We note that  the recently updated version of \cloudy\ was used{}, which incorporates a major revision of the atomic database with more transitions. We included all available atomic levels for low-ionized metals, the large model atoms for the hydrogen and helium iso-sequences, and neglected molecules.

\begin{figure*}[tb]
  \centering
  \includegraphics[draft=false, width=\hsize, trim = 1.2mm 2.5mm 3.5mm 2.5mm, clip]{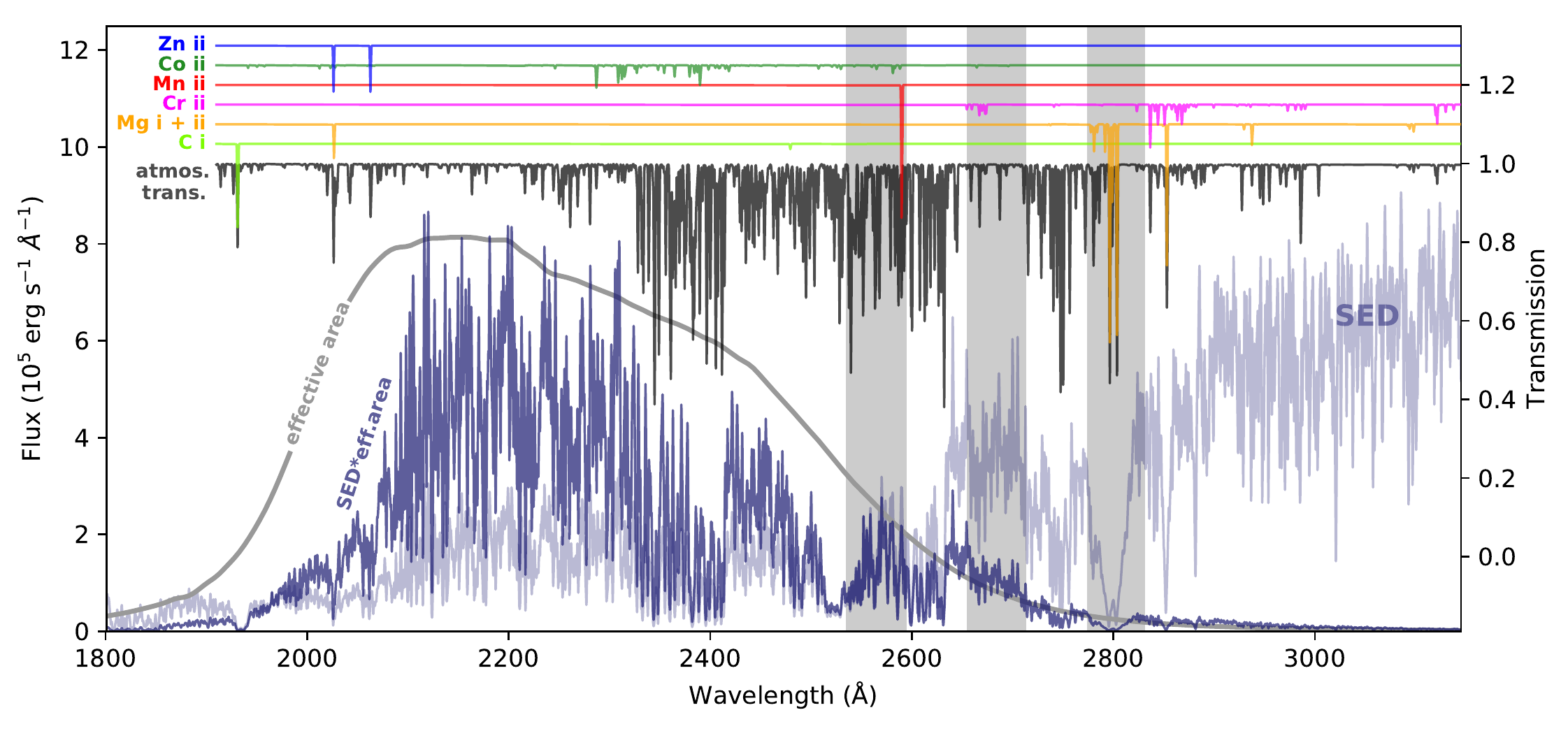}
  \caption{Model of the SED\ of \wtw\  at the distance
           of the planet (light blue). The effective area of the instrument-filter
           combination and its multiplication with the stellar spectrum are depicted
           in arbitrary units.
           An atmospheric transmission model is depicted in black and
           contributions of individual ions are shown above --- all remaining
           lines are caused by \ion{Fe}{ii}.
           Ranges covered by HST COS observations of \wt{} are shaded.
           All spectra are shown at a resolution of 3000, the nominal
           resolution of the CUTE cubesat \citep{Fleming2018}.
           }
  \label{fig:filter}
\end{figure*} 

We setup a gas cloud with a line-of-sight extent of 3~\Rpl{} and use conditions that can exist in the thermosphere of a strongly evaporating hot Jupiter: a uniform gas cloud with solar metallicity, with a temperature of $10^4$~K, and a density of $10^{10}$~\ccm{}, leading to a column density of $10^{20.6}$~cm$^{-2}$.
For the irradiation of the gas, we compute the spectral energy distribution (SED) of \wtw{} with the stellar parameters of \citet{Delrez2016} using the atmosphere code from \citet{Shulyak2004}. The computed stellar SED is supplemented by XUV radiation with the reconstructed irradiation level following \citet[][]{Salz2016}. Figure~\ref{fig:filter} shows the stellar SED  and the observed flux in the \uvmt{} filter.

The simulation exhibits many \ion{Fe}{ii} absorption lines densely distributed over the spectral range covered by the \uvmt{} filter (see Fig.~\ref{fig:filter}). A few mostly singly ionized metals contribute to the absorption, most notably the \ion{Mg}{ii} lines around 2800\AA{}, which were  detected in \wtb{} \citep{Haswell2012}. The \ion{Mg}{i} line at 2852\AA{} was not covered by the \wt{} data, but was seen in absorption in \hdtb{}  \citep{Vidal2013}. The central order of the COS data provided the least evidence for excess absorption by \wtb{}, which is reproduced by our simulation. Relatively strong broad-band absorption extends up to about 3000~\AA{}, at which point absorption lines become more sparse. With the SED of \wtw{} in the \uvmt{} filter, we find 0.5\% broad-band absorption from this model if 15\% of the stellar disk is covered by the cloud. Without the \ion{Fe}{ii} ion the absorption level is a factor ten weaker. The computed absorption level is on a par with what is suggested by the observations, showing that \ion{Fe}{ii} may be responsible for broad-band NUV absorption.

\section{Conclusions}\label{SectConclusion}
%
With an archival transit campaign of \ffce{}, we demonstrate that \swift{} UVOT achieves an accuracy of about 0.5\% for relative flux measurements, which is sufficient for transit observations of hot Jupiters.
With three transit observations of \wtwb{}, we detect the planetary transit in the NUV. Our analysis suggests that the broad-band NUV transit depth of the hot Jupiter is 0.55 percentage points deeper than in the optical, but the significance of this result remains low (1.9$\sigma$). Uncertainties in the out-of-transit level and the instrumental error of the UVOT photometry limit the accuracy with which the NUV transit depth can be measured.

Our \cloudy{} simulations show that broad-band NUV absorption can be caused by \ion{Fe}{ii} in an extended atmosphere. Similar excess NUV absorption is only known in the \wt{} system. Planets with one thousand times lower mass-loss estimates than \wtwb{} and \wtb{} do not exhibit broad-band NUV absorption, providing evidence that such absorption may indeed by linked to the extreme planetary mass loss. Future NUV transit-depth measurements provide the opportunity to undertake systematic studies of escaping planetary atmospheres and to test theoretical predictions like the energy-limited nature of planetary mass loss. With a planned mission start in 2020, the CUTE cubesat will address this subject with spectrally resolved NUV transit light curves of short-period planets orbiting bright stars \citep{Fleming2018}.

\begin{acknowledgements}
We thank Paul Kuin from the Swift UVOT team for the his help regarding the
UVOT data reduction. We would also like to thank the second referee for the insightful comments.
This work made use of data supplied by the UK Swift Science Data Centre at the University of Leicester.
This work has made use of data from the European Space Agency (ESA) mission
{\it Gaia} (\url{https://www.cosmos.esa.int/gaia}), processed by the {\it Gaia}
Data Processing and Analysis Consortium (DPAC,
\url{https://www.cosmos.esa.int/web/gaia/dpac/consortium}). Funding for the DPAC
has been provided by national institutions, in particular the institutions
participating in the {\it Gaia} Multilateral Agreement.
MS acknowledges support by the DFG SCHM~1032/57-1 and DLR 50OR1710.
PCS acknowledges support by DLR 50OR1706.
\end{acknowledgements}

\bibliographystyle{aa}
\setlength{\bibsep}{0.0pt}
\bibliography{wasp121}

\appendix

\section{Low-sensitivity patches of UVOT}\label{Sect:patches}
%
\begin{figure}[h]
  \centering
  \includegraphics[draft=false, width=\hsize, trim = 1.2mm 2.5mm 3.5mm 2.5mm, clip]{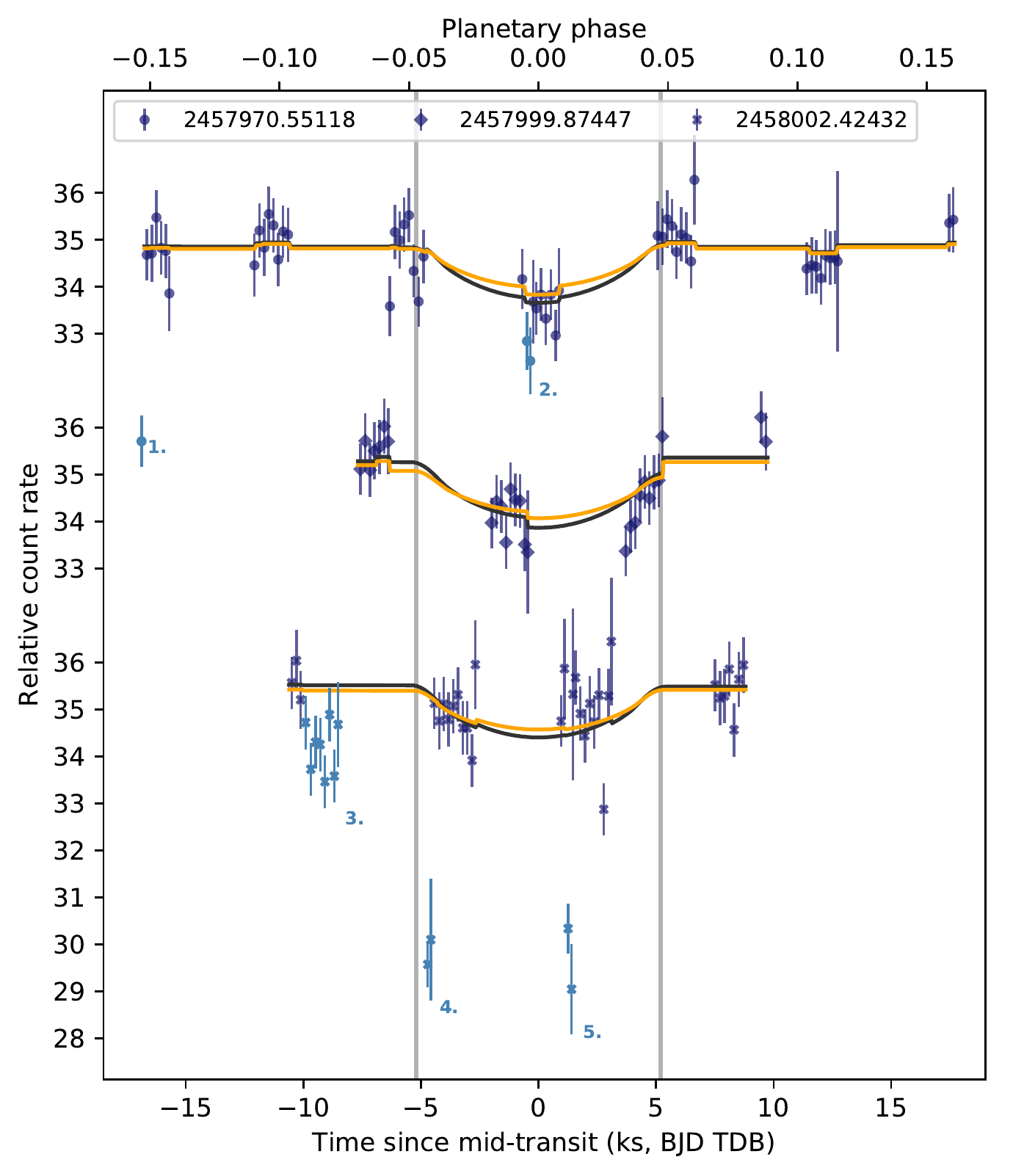}
  \caption{Transit light curves of \wtw{} with our model including
           the offsets caused by instrument systematics (black curve).
           A model with the transit depth fixed to the broad-band optical value
           is shown by the orange line.
           Exposures discarded due to low-sensitivity patches
           are indicated; see labels 1 to 5.
           }
  \label{fig:discard}
\end{figure} 

\begin{figure}[h]
  \centering
  \includegraphics[draft=false, width=\hsize]{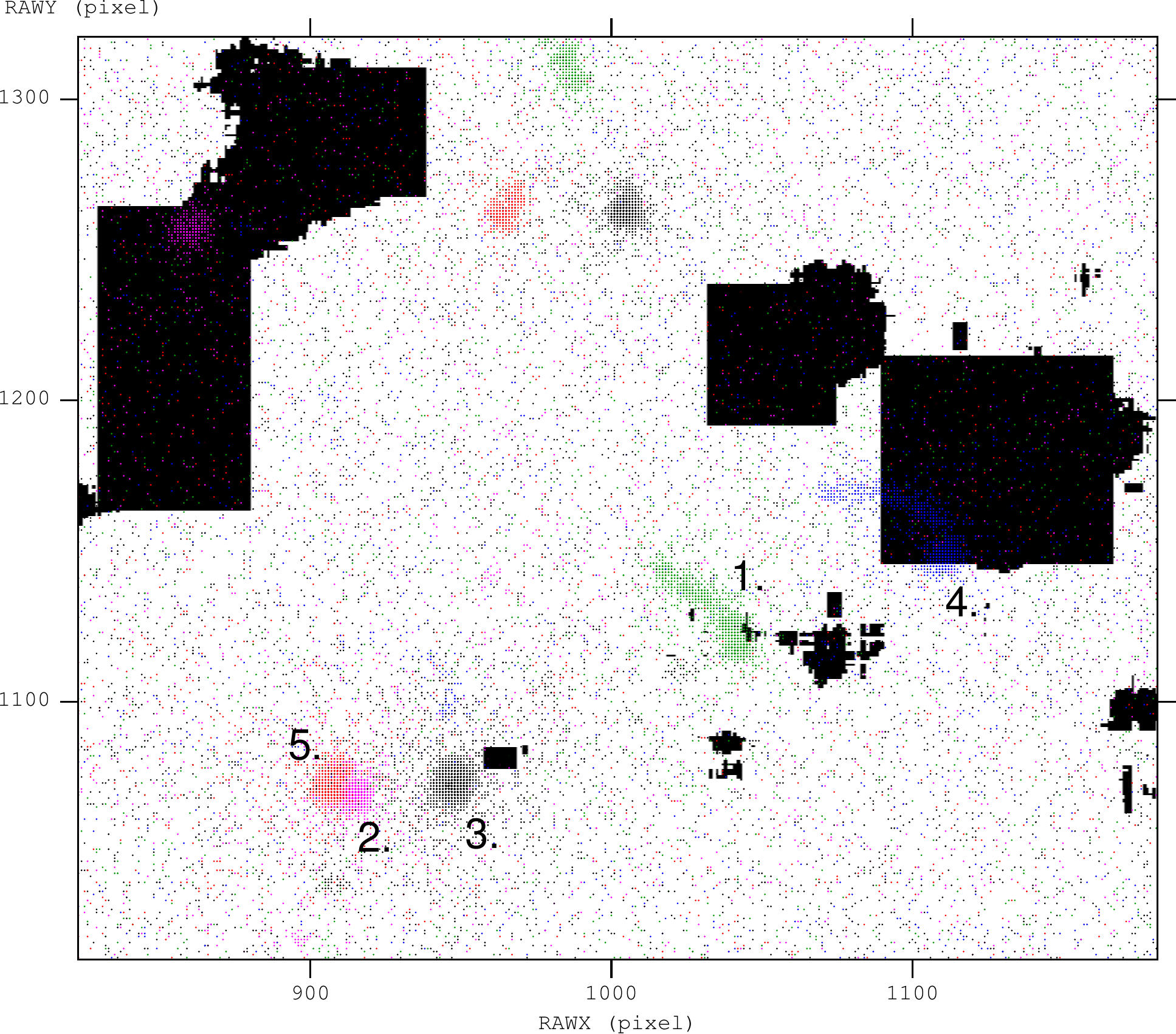}
  \caption{Source positions in raw detector coordinates during exposures affected by
           low-sensitivity patches.
           The figure shows a central region of the UVOT; known bad patches
           are shaded.
           The source positions are numbered according to Fig.~\ref{fig:discard}.
           The aimpoint was close to the XY position 935 and 1060 in raw detector coordinates.
           }
  \label{fig:patches}
\end{figure} 

The UVOT has known patches of low sensitivity. Usually, a double-correction-slew places the source on clean detector regions for the long science exposure. After the first slew onto the target, the spacecraft retains considerable motion over the first acquisition exposure. This causes the distorted PSF during exposures 1 and 4 in Fig.~\ref{fig:patches}.
A correction slew is then performed, after which the target usually falls within 50 pixels of the aimpoint. The second acquisition exposure is obtained, and after another correction slew, we find the target within 18 pixels of the aimpoint. Continuous monitoring for the rest of the visibility window follows (science exposure). 

We identify and exclude four acquisition and one science exposure that are affected by low sensitivity patches (see Fig.\ref{fig:discard} and \ref{fig:patches}). {\bf 1 and 4:} In visit 1 - orbit 1 and visit 3 - orbit 2 the source position in the first acquisition exposure coincides with known low-sensitivity patches. {\bf 3:} During visit 3 - orbit 1 the source position in the science exposure touches a low-sensitivity patch and the source flux is diminished by about 4\% compared to the acquisition exposures in the same orbit. {\bf 5:} In visit 3 - orbit 3 the second acquisition exposure shows a clearly diminished flux compared with the first and third exposures. There is no known low-sensitivity patch at the source position of (907, 1073) in raw detector coordinates. It is unlikely that the host star shows a 15\% drop in flux that happens simultaneously with position changes on the detector. Therefore, we identify a previously unknown patch at this position.

{\bf 2:} We checked all exposures again for the new patch location, and indeed in the second exposure of visit 1 - orbit 4 the stellar PSF partially overlaps with this position. The source flux is also reduced by about 5\% compared with exposures 1 and 3, which supports our identification of a new low-sensitivity patch. These five exposures were excluded from our transit analysis.

\end{document}